\documentclass[reprint,amsmath,amssymb,aip,nofootinbib, citeautoscript,]{revtex4-2}
\pdfoutput=1
\usepackage{titlesec}
\titleformat*{\section}{\normalsize\bfseries}
\titleformat*{\subsection}{\normalsize\itshape}
\titlespacing*{\section} {0pt}{3ex}{1ex}
\titlespacing*{\subsection} {0pt}{2ex}{0ex}
\usepackage{caption}
\usepackage{graphicx}
\usepackage{dcolumn}
\usepackage{natbib}
\usepackage{bm}
\usepackage{amsbsy}

\DeclareUnicodeCharacter{2061}{}
\usepackage{chemist}
\usepackage[version=4]{mhchem}
\makeatletter
\newcommand{\fmarki}{\textdagger}
\newcommand{\fmarkii}{a}
\newcommand{\fmarkiii}{b}
\newcommand{\fmarkiv}{c}
\def\@alph#1{{\ifcase#1\or \fmarki\or \fmarkii\or \fmarkiii\or \fmarkiv\or \fmarkv\or \fmarkvi\or \fmarkvii\or \fmarkviii\or \fmarkix \else\@ctrerr\fi}}
\makeatother

\makeatletter
\newcommand{\smallsym}[2]{#1{\mathpalette\make@small@sym{#2}}}
\newcommand{\make@small@sym}[2]{%
  \vcenter{\hbox{$\m@th\downgrade@style#1#2$}}%
}
\newcommand{\downgrade@style}[1]{%
  \ifx#1\displaystyle\scriptstyle\else
    \ifx#1\textstyle\scriptstyle\else
      \scriptscriptstyle
  \fi\fi
}
\makeatother

\begin{document}

\title{Nanocrystal Geometry Governs Phase Transformation Pathways \\ in Palladium Hydride}

\author{Daewon Lee}
 \thanks{These authors contributed equally to this work.}
 \affiliation{\mbox{Materials Sciences Division, Lawrence Berkeley National Laboratory, Berkeley, CA 94720, USA}}
 \affiliation{\mbox{Department of Materials Science and Engineering, University of California, Berkeley, CA 94720, USA}}
\author{Sam Oaks-Leaf}
 \thanks{These authors contributed equally to this work.}
 \affiliation{\mbox{Department of Chemistry, University of California, Berkeley, CA 94720, USA}}
\author{Hyeonjong Ma}
 \thanks{These authors contributed equally to this work.}
 \affiliation{\mbox{Materials Sciences Division, Lawrence Berkeley National Laboratory, Berkeley, CA 94720, USA}}
 \affiliation{Department of Energy Science and Engineering, Daegu Gyeongbuk Institute of Science and Technology, Daegu 42988, Republic of Korea}
\author{Jianlong He}
 \affiliation{\mbox{School of Chemistry and Biochemistry, Georgia Institute of Technology, Atlanta, GA 30332, USA}}
\author{Zhiqi Wang}
 \affiliation{\mbox{School of Chemistry and Biochemistry, Georgia Institute of Technology, Atlanta, GA 30332, USA}}
\author{Yifeng Shi}
 \affiliation{\mbox{School of Chemical and Biomolecular Engineering, Georgia Institute of Technology, Atlanta, GA 30332, USA}}
\author{Eonhyoung Ahn}
 \affiliation{Department of Energy Science and Engineering, Daegu Gyeongbuk Institute of Science and Technology, Daegu 42988, Republic of Korea}
\author{Karen C. Bustillo}
 \affiliation{National Center for Electron Microscopy, Molecular Foundry, Lawrence Berkeley National Laboratory, Berkeley, CA 94720, USA}
\author{Chengyu Song}
 \affiliation{National Center for Electron Microscopy, Molecular Foundry, Lawrence Berkeley National Laboratory, Berkeley, CA 94720, USA}
\author{Stephanie M. Ribet}
 \affiliation{National Center for Electron Microscopy, Molecular Foundry, Lawrence Berkeley National Laboratory, Berkeley, CA 94720, USA}
 \author{Rohan Dhall}
\affiliation{National Center for Electron Microscopy, Molecular Foundry, Lawrence Berkeley National Laboratory, Berkeley, CA 94720, USA}
\author{Colin Ophus}
 \affiliation{\mbox{Department of Materials Science and Engineering, Stanford University, Stanford, CA 94305, USA}}
\author{Mark Asta}
 \affiliation{\mbox{Materials Sciences Division, Lawrence Berkeley National Laboratory, Berkeley, CA 94720, USA}}
 \affiliation{\mbox{Department of Materials Science and Engineering, University of California, Berkeley, CA 94720, USA}}
\author{Jiwoong Yang}
 \affiliation{Department of Energy Science and Engineering, Daegu Gyeongbuk Institute of Science and Technology, Daegu 42988, Republic of Korea}
  \affiliation{Energy Science and Engineering Research Center, Daegu Gyeongbuk Institute of Science and Technology, Daegu 42988, Republic of Korea}
\author{Younan Xia\textsuperscript{*}}
 \email{younan.xia@bme.gatech.edu}
 \affiliation{\mbox{School of Chemistry and Biochemistry, Georgia Institute of Technology, Atlanta, GA 30332, USA}}
 \affiliation{The Wallace H. Coulter Department of Biomedical Engineering, Georgia Institute of Technology and Emory University, Atlanta, GA 30332, USA}
\author{David T. Limmer\textsuperscript{*}}
 \email{dlimmer@berkeley.edu}
 \affiliation{\mbox{Materials Sciences Division, Lawrence Berkeley National Laboratory, Berkeley, CA 94720, USA}}
 \affiliation{\mbox{Department of Chemistry, University of California, Berkeley, CA 94720, USA}}
 \affiliation{\mbox{Chemical Sciences Division, Lawrence Berkeley National Laboratory, Berkeley, CA 94720, USA}}
 \affiliation{\mbox{Kavli Energy Nanoscience Institute, Berkeley, CA 94720, USA}}
\author{Haimei Zheng\textsuperscript{*}}
 \email{hmzheng@lbl.gov} 
 \affiliation{\mbox{Materials Sciences Division, Lawrence Berkeley National Laboratory, Berkeley, CA 94720, USA}}
 \affiliation{\mbox{Department of Materials Science and Engineering, University of California, Berkeley, CA 94720, USA}}

\date{\today}

\begin{abstract}
Pathways and structural dynamics of phase transformations impact performance of materials in energy and information storage technologies. Palladium hydride ($\mathrm{PdH}_x$) nanocrystals are an ideal model system for studying solute-induced phase transformations, where elastic energy from lattice mismatch between $\alpha$-$\mathrm{PdH}_x$ and $\beta$-$\mathrm{PdH}_x$ phases is often considered a key to determining the transformation pathways. $\alpha$/$\beta$-$\mathrm{PdH}_x$ interfacial elastic energy is affected by the confined geometry of a nanocrystal. However, how nanocrystal geometry influences phase transformation pathways is largely unknown. Using \textit{in situ} liquid phase transmission electron microscopy, we directly visualize hydrogenation in Pd nanocrystals with two geometries—a nanocube and a hexagonal nanoplate. Both follow similar sequences of an initially curved nucleus, interface flattening, and reverse-stage nucleation; however, their evolving $\alpha$/$\beta$-$\mathrm{PdH}_x$ interfaces exhibit geometry-dependent crystallographic alignments. In nanocubes, \{100\}-aligned configurations conform to static elastic energy ordering, representing a pathway that maintains a local mechanical equilibrium, whereas nanoplates display both \{110\}- and \{211\}-aligned interfaces. Theoretical simulations show that geometry determines the accessibility of alternative phase transformation pathways as the system is driven far from equilibrium during hydrogenation. These findings identify geometry as a fundamental parameter for directing phase transformation pathways, offering design principles for accessing atypical configurations and improving properties of intercalation-based devices.
\end{abstract}

\maketitle

\section*{\label{sec:intro}Introduction}

Phase transformations driven by solute incorporation are pervasive in solid materials and lie at the core of numerous technologies, ranging from hydrogen storage~\cite{pundt_hydrogen_2004, griessen_thermodynamics_2016} to lithium-ion batteries~\cite{cogswell_theory_2013,sood_electrochemical_2021}, nanocrystal synthesis~\cite{son_cation_2004,rivest_cation_2013}, non-volatile memories~\cite{lu_electric_2017,huang_electrolyte_2019}, and neuromorphic computing~\cite{zhu_ionic_2019,nadkarni_modeling_2019}. During such transformations, the accommodation of solute species is intimately coupled with atomic rearrangements in the host material, leading to intricate structural dynamics that are fundamental to material performance. Understanding and controlling these coupled processes are thus essential for advancing energy and information storage technologies that rely on solute-mediated phase transformations.

Palladium hydride ($\mathrm{PdH}_x$) serves as a model system for investigating solute-induced phase transformations~\cite{baldi_in_situ_2014,ulvestad_self_2017}, owing to the ability of palladium to readily absorb and desorb hydrogen under practically achievable temperatures and pressures~\cite{baldi_in_situ_2014,wicke_hydrogen_1978,dekura_hydrogen_2019}. The $\mathrm{PdH}_x$ system features two compositionally distinct, immiscible phases—hydrogen-poor $\alpha$-$\mathrm{PdH}_x$ and hydrogen-rich $\beta$-$\mathrm{PdH}_x$ phases. Both of these phases have the face-centered cubic (\textit{fcc}) crystal structure. Due to the co-existence of these two phases, hydrogen-induced phase transformations in $\mathrm{PdH}_x$ generally exhibit pronounced spatial heterogeneity. Such spatial heterogeneity requires direct visualization of phase-morphology evolution to elucidate the underlying principles of these transformations.

Furthermore, the lattice mismatch of $\sim$3.5\% between the $\alpha$ and $\beta$ phases~\cite{wicke_hydrogen_1978,wyckoff_structure_1924} generates substantial coherency strain at the $\alpha$/$\beta$-$\mathrm{PdH}_x$ interfaces~\cite{syrenova_hydride_2015,ulvestad_three_2017}. Elastic energy at these interfaces thus has often been invoked to explain the observed hydrogen intercalation pathways in Pd nanocrystals~\cite{ulvestad_three_2017,narayan_direct_2017,angell_lattice_2022,lee_atomic_2024}; namely, the evolving morphology and propagation of phase boundaries during hydrogen-induced transformations. Because the confined geometry of a nanocrystal influences $\alpha$/$\beta$-$\mathrm{PdH}_x$ interfacial elastic energy~\cite{ulvestad_three_2017,lee_atomic_2024}, systematic variations in nanocrystal geometry could, in principle, regulate which phase transformation pathways are accessible. Despite this possibility, the role of geometry in enabling alternative transformation pathways has not been clearly established.

Here, we examine Pd nanocrystals with two geometries—nanocubes and hexagonal nanoplates—to investigate how geometry governs hydrogen-induced phase transformation pathways. Using liquid phase transmission electron microscopy (LP-TEM), where hydrogen is generated through radiolysis of water in the encapsulated aqueous solution, we directly visualize the propagation of $\beta$-$\mathrm{PdH}_x$ within individual $\alpha$-$\mathrm{PdH}_x$ nanocrystals. Both geometries undergo similar transformation steps—an initially curved nucleus, interface flattening at intermediate stages, and a reverse nucleation stage—yet their $\alpha$/$\beta$-$\mathrm{PdH}_x$ interfaces adopt distinct crystallographic alignments. At intermediate stages, nanocubes develop \{100\}-aligned interfaces, consistent with elastic energy ordering, whereas nanoplates display both \{110\}- and \{211\}-aligned configurations. At comparable $\alpha$/$\beta$ phase fractions, the elastic interfacial energies of these two crystallographic alignments in the nanoplates differ substantially; thus, the observed hydrogenation pathways cannot be explained solely by static energy minimization. Theoretical analyses reveal that the nanocrystal geometry dictates which phase transformation pathways become accessible as the system is pushed far from equilibrium during the hydrogenation process. While the nanoplate geometry enables an alternative, qualitatively distinct pathway, nanocubes remain confined to a transformation route that conforms to static energetics under the identical chemical driving force.

\section*{\label{sec:results}Results and Discussion}

\subsection*{\label{sec:struct}Structural characteristics of model Pd nanocrystal systems}

\textbf{Figure 1} presents the structural characteristics of Pd nanocubes and hexagonal nanoplates (hereafter referred to as nanoplates), which serve as model systems to investigate how nanocrystal geometry governs hydrogen-induced phase transformation pathways. The schematic illustration in \textbf{Figure 1a} depicts hydrogenation of nanocubes and nanoplates, where intercalated hydrogen atoms occupy interstitial sites~\cite{dekura_electronic_2018,schneemann_nanostructured_2018} of the \textit{fcc} Pd lattice. The Pd nanocubes and nanoplates are synthesized using our established protocols~\cite{xiong_polyvinyl_2006,jin_synthesis_2011}. In this study, the nanocubes exhibit edge lengths of $\sim$19–23 nm, and the nanoplates have edge lengths of $\sim$30–70 nm and a thickness of $\sim$20–23 nm. The thickness is independently confirmed by side-view high-resolution TEM (HRTEM) imaging and atomic force microscopy (AFM) topographic mapping (\textbf{Figure S1}).

We employ scanning TEM (STEM) and TEM techniques to describe and compare the geometries of the two model nanocrystals in terms of crystallographic directions and planes (\textbf{Figure 1b} and \textbf{1c}). Atomic-resolution HRSTEM imaging (see also \textbf{Methods} and \textbf{Figure S2} regarding post-acquisition drift correction) reveals that nanocubes possess a square lattice with four-fold rotational symmetry, consistent with \{100\} facet termination, whereas nanoplates exhibit a hexagonal lattice with six-fold rotational symmetry, suggesting \{111\} top and bottom facets. Side-view HRTEM imaging and fast Fourier transform (FFT) patterns (\textbf{Figure S3}) further confirm the \{111\} termination of the top and bottom surfaces, and also show that the side facets of nanoplates are bound by a combination of \{100\} and \{111\} planes. STEM images and corresponding energy-dispersive X-ray spectroscopy (EDS) elemental maps verify that the synthesized nanocrystals are composed of Pd (\textbf{Figure S4}).

Furthermore, in \textbf{Figure 1b} and \textbf{1c}, HRTEM-based FFT and selected area electron diffraction (SAED) analyses (see \textbf{Figure S5} for the selected area aperture) corroborate the HRSTEM and side-view HRTEM results. Nanocubes present \{200\} and \{220\} reflections along the [00-1] zone axis, whereas nanoplates show \{220\} reflections together with forbidden $\frac{1}{3}$\{422\} reflections along the [11-1] zone axis. The $\frac{1}{3}$\{422\} reflections in nanoplates~\cite{xiong_polyvinyl_2006,kirkland_structural_1993,xiong_kinetically_2005,germain_stacking_2003} arise from planar defects such as stacking faults lying parallel to \{111\} planes (see \textbf{Figure S3}).

We highlight that Pd nanocubes and hexagonal nanoplates are deliberately chosen as model systems because their projected geometries allow straightforward identification of crystallographic directions and facets in S/TEM images. For nanocubes, the sides of the projected rectangle are perpendicular to $\langle100\rangle$ directions, while the $\langle110\rangle$ directions lie at $45 ^\circ$ relative to the sides. For hexagonal nanoplates, the edges of the projected hexagon are perpendicular to $\langle211\rangle$ directions, while the tangents at the vertices are perpendicular to $\langle110\rangle$ directions. This direct geometric correspondence provides an intuitive framework for elucidating $\alpha$/$\beta$-$\mathrm{PdH}_x$ interface alignments throughout phase transformations.

\begin{figure*}[t]

\centering

\includegraphics[width=0.8\textwidth]{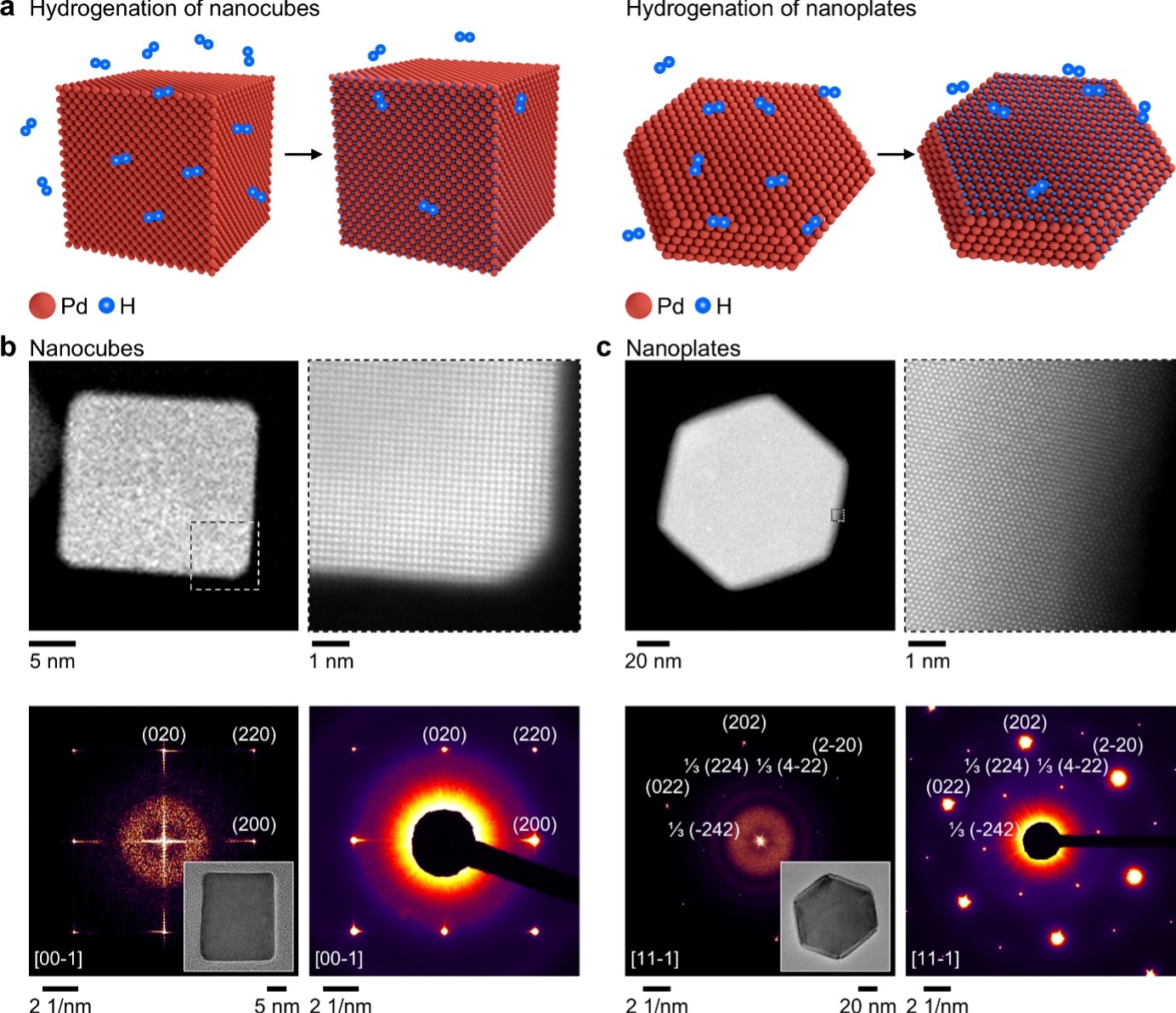}

\caption{Pd nanocrystals as model systems to reveal how nanocrystal geometry governs hydrogen-induced phase transformation pathways. \textbf{a}, Schematic illustration of hydrogenation of Pd nanocubes (left) and nanoplates (right). \textbf{b},\textbf{c}, Representative (HR)STEM, HRTEM, and SAED analyses of Pd nanocubes (\textbf{b}, along the [00-1] zone axis) and nanoplates (\textbf{c}, along the [11-1] zone axis) used in this study. High-resolution STEM images, obtained by enlarging boxed regions from the same nanocrystals shown in the low-magnification STEM images, resolve atomic structures. For separately selected, crystallographically equivalent nanocrystals, FFT patterns derived from HRTEM images (insets) and SAED patterns reveal consistent features: nanocubes exhibit \{200\} and \{220\} reflections, whereas nanoplates display \{220\} reflections together with forbidden $\frac{1}{3}$\{422\} reflections.}

\label{fig:1}

\end{figure*}

\subsection*{In situ LP-TEM observation of nanocrystal hydrogenation}

\begin{figure*}[t]

\centering

\includegraphics[width=\textwidth]{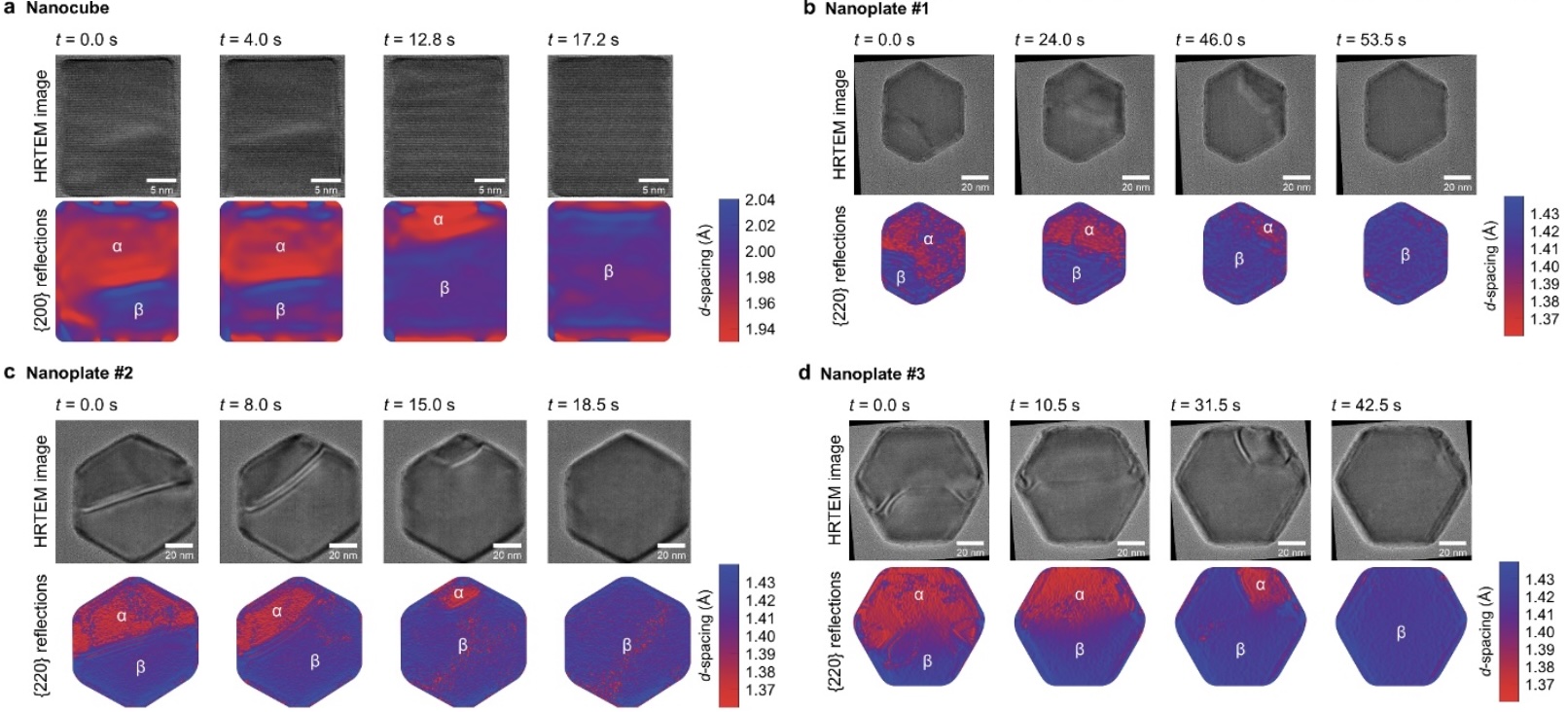}

\caption{\textit{In situ} liquid phase TEM observation of the propagation of the hydrogen-rich $\beta$-$\mathrm{PdH}_x$ phase into the hydrogen-poor $\alpha$-$\mathrm{PdH}_x$ nanocube and nanoplates during hydriding reactions. \textbf{a}–\textbf{d}, Sequential \textit{in situ} HRTEM images from \textbf{Movies S1–S4} (top) and corresponding amplitude-weighted \textit{d}-spacing colormaps (bottom) showing the temporal evolution of $\alpha$- and $\beta$-$\mathrm{PdH}_x$ phase regions during hydrogenation of Pd Nanocube (\textbf{a}), Nanoplate \#1 (\textbf{b}), Nanoplate \#2 (c), and Nanoplate \#3 (\textbf{d}). The \textit{d}-spacing maps are generated using the \{200\} reflections for the Nanocube and the \{220\} reflections for Nanoplates. See \textbf{Figure S7} for amplitude-weighted $\frac{1}{3}$\{422\} \textit{d}-spacing maps for Nanoplates. The hydrogen-induced phase transformation proceeds through four distinct stages: an initial nucleation event, a middle stage characterized by relatively flat $\alpha$/$\beta$-$\mathrm{PdH}_x$ interphase boundaries, a later stage resembling the reverse counterpart of the initial nucleation, and a fully transformed state. Corresponding FFT patterns of the HRTEM images are provided in \textbf{Figure S8} and \textbf{S9}.}

\label{fig:2}

\end{figure*}

Having established these structural characteristics, we perform \textit{in situ} LP-TEM hydrogenation experiments on Pd nanocubes and nanoplates using carbon-film-based liquid cells~\cite{lee_atomic_2024,zhang_defect_2022} (see \textbf{Methods} and \textbf{Figure S6}). The liquid cells are assembled by sandwiching two TEM grids coated with ultrathin carbon films (3–4 nm thick). This configuration encapsulates Pd nanocrystals and a thin film of aqueous KOH solution, held together by van der Waals forces between the carbon films. By exploiting electron-beam irradiation, we induce radiolysis of the encapsulated aqueous solution, which generates molecular hydrogen as the dominant product ~\cite{lee_atomic_2024,hart_hydrated_1964,grogan_bubble_2014,schneider_electron_2014} and initiates hydrogenation of the Pd nanocrystals.

Our LP-TEM technique facilitates \textit{in situ} HRTEM imaging of Pd nanocrystals undergoing hydriding reactions; observation of these reactions has been highly challenging with common gas-phase TEM~\cite{angell_lattice_2022,lee_atomic_2024}. Because the phase transformation from hydrogen-poor $\alpha$-$\mathrm{PdH}_x$ to hydrogen-rich $\beta$-$\mathrm{PdH}_x$ involves a $\sim$3.5\% increase in lattice constant~\cite{wicke_hydrogen_1978,wyckoff_structure_1924}, HRTEM imaging enables quantitative tracking of the hydrogen-induced process. Specifically, we employ geometric phase analysis (GPA) ~\cite{hytch_analysis_1997,hytch_quantitative_1998}, selecting Bragg reflections of interest in the Fourier domain of HRTEM images, to generate \textit{d}-spacing colormaps that delineate expanded ($\beta$-$\mathrm{PdH}_x$) and unexpanded ($\alpha$-$\mathrm{PdH}_x$) lattice regions within individual nanocrystals.

Time-series HRTEM images combined with these colormaps visualize the propagation of $\beta$-$\mathrm{PdH}_x$ within an $\alpha$-$\mathrm{PdH}_x$ nanocube (“Nanocube”) and nanoplates (“Nanoplate \#1”, “Nanoplate \#2”, and “Nanoplate \#3”) during hydrogen absorption (\textbf{Figure 2} and \textbf{S7}). The Bragg reflections used to generate the \textit{d}-spacing colormaps are annotated on the corresponding FFT patterns obtained from the HRTEM image sequences (\textbf{Figure S8} and \textbf{S9}). To account for all reflections within each crystallographic family, we combine the \textit{d}-spacing information from individual reflections by weighting the squared amplitude of the respective Bragg-filter outputs (see \textbf{Methods} and \textbf{Figure S10}). Specifically, the \{200\} maps are constructed from the (020) and (200) reflections, the \{220\} maps from the (022), (202), and (2-20) reflections, and the $\frac{1}{3}$\{422\} maps from $\frac{1}{3}$(-242), $\frac{1}{3}$(224), and $\frac{1}{3}$(4-22).

We show that, despite their distinct morphologies, both the nanocube and the nanoplates undergo hydriding through similar evolution steps of $\alpha$- and $\beta$-$\mathrm{PdH}_x$ phase regions: an initially curved nucleus of the $\beta$-$\mathrm{PdH}_x$ phase; a middle stage—when approximately half of the $\alpha$ phase is converted into $\beta$—characterized by relatively flat $\alpha$/$\beta$-$\mathrm{PdH}_x$ interfaces; a later stage resembling the reverse counterpart of the initial nucleation; and finally a fully transformed $\beta$-phase state. As confirmed by our theoretical modeling, in both geometries, the interface of the initial nucleus is energetically favored to curve between adjacent sides of the nanocrystal before propagating inhomogeneously to form an energetically favorable, flat interface at near-equal $\alpha$/$\beta$ phase fractions. It is also worthwhile to note that, in the nanoplates, amplitude-weighted \textit{d}-spacing colormaps generated from both \{220\} and defect-induced $\frac{1}{3}$\{422\} reflections demonstrate qualitatively identical evolution of $\alpha$- and $\beta$-$\mathrm{PdH}_x$ phase distributions (\textbf{Figure 2} and \textbf{S7}). This consistency can be explained by planar defects extending across the nanoplate parallel to \{111\} planes (\textbf{Figure S3}).

\subsection*{Structural evolution of $\alpha$/$\beta$-$\mathrm{PdH}_x$ interfaces in different geometries}

\begin{figure*}[t]

\centering

\includegraphics[width=0.8\textwidth]{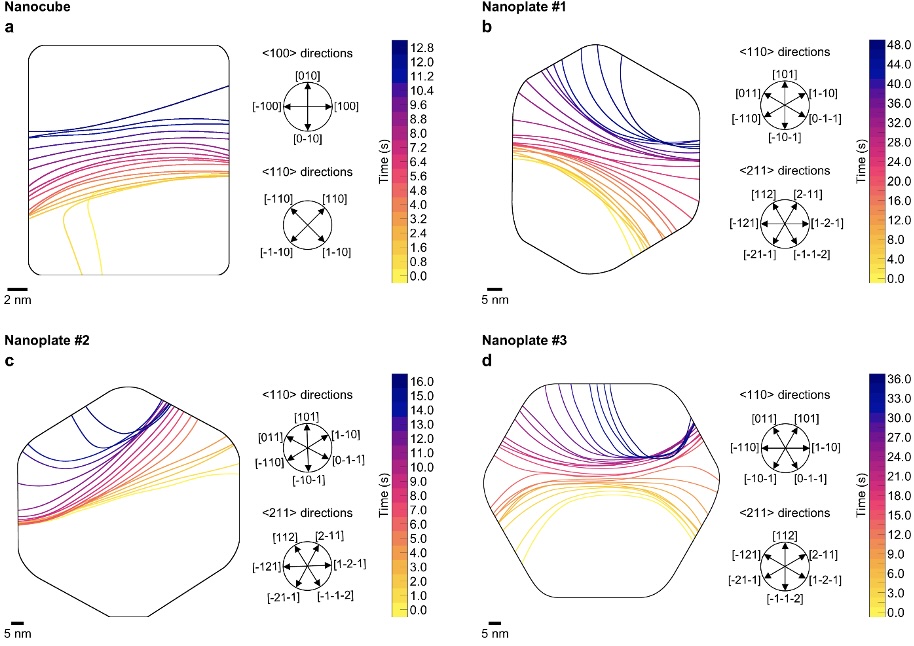}

\caption{Temporal evolution of propagating $\alpha$/$\beta$-$\mathrm{PdH}_x$ interfaces during hydrogenation of Pd nanocube and nanoplates. \textbf{a}–\textbf{d}, Contour plots showing the evolution of $\alpha$/$\beta$-$\mathrm{PdH}_x$ interfaces in the $\mathrm{PdH}_x$ nanocrystals presented in \textbf{Figure 2}: Nanocube (\textbf{a}), Nanoplate \#1 (\textbf{b}), Nanoplate \#2 (\textbf{c}), and Nanoplate \#3 (\textbf{d}). The contours represent $\alpha$/$\beta$-$\mathrm{PdH}_x$ interfaces manually traced from amplitude-weighted \{200\}, \{220\}, and $\frac{1}{3}$\{422\} \textit{d}-spacing colormaps corresponding to sequential HRTEM images. Representative sequential image series used for contour extraction are provided in \textbf{Figure S11–S14}. The crystallographic directions indicated by schematic diagrams with arrows are assigned based on the FFT patterns in \textbf{Figure S8} and \textbf{S9}, using the [00-1] zone axis for the nanocube and the [11-1] zone axis for the nanoplates as guidance.}

\label{fig:3}

\end{figure*}

Although the overall transformation steps are similar, the crystallographic directions and related planes predominantly associated with $\alpha$/$\beta$-$\mathrm{PdH}_x$ interface alignments are nanocrystal-geometry-dependent. In nanocubes, the $\alpha$/$\beta$ interphase boundary orientations are most clearly described along the $\langle100\rangle$ and/or $\langle 110 \rangle$ directions~\cite{lee_atomic_2024}. By contrast, in nanoplates, the interface alignments are more suitably represented along the $\langle 110 \rangle$ and/or $\langle 211 \rangle$ directions. These directional sets, as identified from FFT and SAED patterns (\textbf{Figure 1b}, \textbf{1c}, \textbf{S8}, and \textbf{S9}), provide a basis for describing the structural evolution of the propagating $\alpha$/$\beta$-$\mathrm{PdH}_x$ interface.

Our GPA-based \textit{d}-spacing mapping method enables detailed extraction of propagating $\alpha$/$\beta$-$\mathrm{PdH}_x$ interphase boundaries as contour lines from time-series HRTEM images during hydriding reactions (\textbf{Figure 3} and \textbf{S11–S14}). In nanoplates, since \textit{d}-spacing maps from both \{220\} and forbidden $\frac{1}{3}$\{422\} reflections show similar phase morphology evolution, one representative contour is used to track the $\alpha$/$\beta$ phase boundary shape in each HRTEM image, as demonstrated for Nanoplate \#1–\#3 (\textbf{Figure S12–S14}).

In the Nanocube (\textbf{Figure 3a} and \textbf{S11}), the advancing $\alpha$/$\beta$-$\mathrm{PdH}_x$ interface exhibits a curved–flat–curved evolution: it begins with a curved profile at the initial stage (0.0–0.8 s), develops into a \{100\}-aligned configuration at an intermediate stage (exemplified at 10.4 s), and eventually returns to a curved profile at the late stage (12.8 s). This intermediate flattening into \{100\} alignment is consistent with previous reports on the hydrogenation of \{100\}-faceted Pd nanocubes~\cite{narayan_direct_2017,lee_atomic_2024,sytwu_visualizing_2018}. Such behavior further conforms to elastic energy ordering of facets, as \{100\} represents the lowest-energy $\alpha$/$\beta$-$\mathrm{PdH}_x$ interface when the $\alpha$- and $\beta$-phase fractions are approximately comparable in the nanocube geometry~\cite{lee_atomic_2024} (\textbf{Figure S15}).

By contrast, at intermediate stages, the nanoplates reveal $\alpha$/$\beta$-$\mathrm{PdH}_x$ interface alignments along different crystallographic planes (\textbf{Figure 3b–3d} and \textbf{S12-14}). For Nanoplate \#1 (\textbf{Figure 3b} and \textbf{S12}), the interphase boundary develops into a moderately flat profile at the intermediate stage (28.0 s), appearing aligned with \{110\}, whereas the early (e.g., 0.0–4.0 s) and late stages (e.g., 44.0–48.0 s) exhibit curved boundaries. In Nanoplate \#2 (\textbf{Figure 3c} and \textbf{S13}), the phase transformation is captured near an intermediate stage; the $\alpha$/$\beta$-$\mathrm{PdH}_x$ interface at 0.0–1.0 s appears relatively flat, with no clear distinction between \{110\} and \{211\} planes, although its orientation is slightly closer to \{110\}. The interface subsequently evolves into a curved boundary near the nanoplate corner at later times (e.g., 14.0–16.0 s). Nanoplate \#3 (\textbf{Figure 3d} and \textbf{S14}) also displays a curved–flat–curved evolution, with the interface flattening at the intermediate stage (10.5 s), approximately aligned with \{211\} planes. Collectively, the nanoplates exhibit this three-stage interface profile evolution as well, but unlike the nanocubes, the intermediate orientations of the aligned $\alpha$/$\beta$-$\mathrm{PdH}_x$ interface vary with respect to crystallographic planes.

These observations highlight that nanocrystal geometry not only influences the crystallographic orientations involved in phase transformation but also dictates whether facet-dependent elastic energy considerations alone suffice to describe the structural evolution of the $\alpha$/$\beta$-$\mathrm{PdH}_x$ phase boundary. In nanocubes, the \{100\} alignment at the intermediate stage conforms well to elastic energy ordering (\textbf{Figure S15}). In contrast, the nanoplates display the $\alpha$/$\beta$-$\mathrm{PdH}_x$ interface alignments along both \{110\} and \{211\} during the mid-transformation stage. As demonstrated in our later calculations, \{211\} does not yield an elastic energetic advantage over \{110\} at the stage in which the nanoplate contains nearly balanced volume fractions of the $\alpha$ and $\beta$ phases (\textbf{Figure S15}). Thus, its occurrence cannot be rationalized by static energetics alone. This indicates that, in addition to energetics, dynamic contributions must also be considered to fully account for the experimentally observed hydrogen-induced phase transformation pathways.

\subsection*{Theoretical modeling of phase transformation pathways in nanoplates}

\begin{figure*}[t]

\centering

\includegraphics[width=0.70\textwidth]{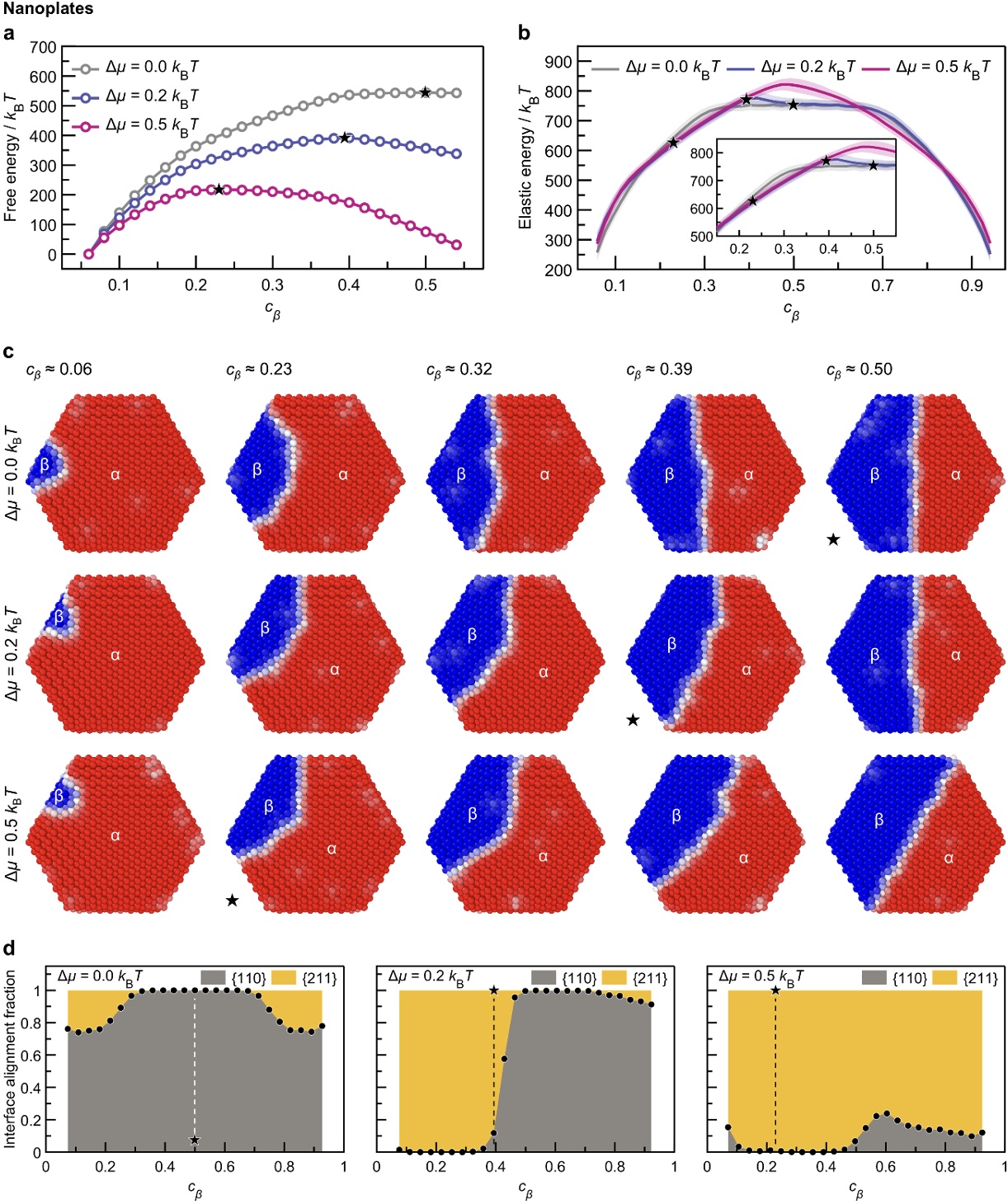}

\caption{Free energetics and simulated hydrogenation dynamics in nanoplates. \textbf{a}, Helmholtz free energy as a function of the $\beta$-phase fraction ($c_\beta$) in the nanoplate. Black pentagram markers denote the transition state locations for each value of the effective chemical driving, $\Delta$$\mu$. \textbf{b}, Average elastic energy across simulated hydrogenation trajectories, plotted as a function of $c_\beta$. The shaded region represents the standard deviation across trajectories. Black pentagram markers correspond to the same transition-state $c_\beta$ values shown in panel \textbf{a} for each curve. All energies are measured in units of $k_\mathrm{B}T$, where $k_\mathrm{B}$ is the Boltzmann constant and $T$ is the simulation temperature. \textbf{c}, Representative snapshots from absorption trajectories in nanoplates. Each row corresponds to a single trajectory at $\Delta$$\mu$ = 0.0, 0.2, and 0.5 $k_\mathrm{B}T$ (top to bottom). Snapshots are colored in terms of the density of each phase, with blue as $\beta$-$\mathrm{PdH}_x$ and red as $\alpha$-$\mathrm{PdH}_x$. Black pentagram markers additionally indicate the snapshots corresponding to the transition state. \textbf{d}, Fraction of $\alpha$/$\beta$-$\mathrm{PdH}_x$ interfaces aligned with either the \{110\} or \{211\} orientations across the ensemble of simulated absorption trajectories. At each value of $c_\beta$, the $\alpha$/$\beta$ interface is computed for all trajectories. The color represents the fraction of best-fit planes whose normal vector is closer to that crystallographic orientation than to the alternative orientation considered. Black pentagram markers and their corresponding vertical lines denote the transition-state $c_\beta$ values.}

\label{fig:4}

\end{figure*}
To gain further insight into how nanocrystal geometry governs the accessible hydrogenation pathways, we employ a model of solid-solid phase transformations, previously parameterized for the $\mathrm{PdH}_x$ system~\cite{lee_atomic_2024}. We consider sites corresponding to $\mathrm{PdH}_x$ unit cells, which are assigned to either the $\alpha$ or $\beta$ phase. The Hamiltonian of the model, based on the elastic Ising model~\cite{fratzl_modeling_1999,fratzl_modeling_1999,frechette_consequences_2019, frechette_elastic_2021}, incorporates the influence of nanocrystal geometry on the elastic energy associated with $\alpha$/$\beta$-$\mathrm{PdH}_x$ co-existence through an effective interaction potential. Dynamics are propagated via kinetic Monte Carlo with rates that obey microscopic reversibility with respect to this potential (see \textbf{Methods}).

This model has been shown to accurately reproduce the bulk critical temperature of $\mathrm{PdH}_x$, as well as its suppression as nanocrystal size decreases. Furthermore, it has been demonstrated that the observed hydrogen sorption pathways in Pd nanocubes could be explained by orienting the $\alpha$/$\beta$-$\mathrm{PdH}_x$ interface to continually minimize the elastic energy cost predicted by the model~\cite{lee_atomic_2024}. Here, we focus on how the preferred interface alignment changes with nanocrystal geometry and on the propensity of each nanocrystal system to deviate from the reversible pathway.

We first recreate an idealized version of the nanoplate geometry observed in experiment: a regular hexagon consisting of 2,133 unit cells and truncated by \{111\} and \{100\} facets. To establish the relevant thermodynamic properties, we calculate the Helmholtz free energy as a function of the $\beta$-phase fraction (\textbf{Figure 4a}). Then, we observe how the free energy changes in response to the chemical driving of hydrogenation by reweighting those free energies with an added term of -$\Delta$$\mu$N$c_\beta$. Here, $\Delta$$\mu$ is the chemical potential difference between hydrogen in solution and in the $\beta$-$\mathrm{PdH}_x$ phase, $N$ is the number of unit cells in the system, and $c_\beta$ is the $\beta$-phase fraction. In particular, we can identify the degree to which the free energy maximum (i.e., the transition state) shifts to lower values of $c_\beta$ in response to a specific chemical driving strength.

While free energy sampling gives us access to the thermodynamically preferred configurations, a hydrogen absorption process occurring over a finite time may drive the $\alpha$/$\beta$-$\mathrm{PdH}_x$ interface away from the free-energy-minimized path. We therefore use kinetic Monte Carlo to generate an ensemble of trajectories from the $\alpha$ to the $\beta$ phase at each value of chemical driving (see \textbf{Methods}). The simulated hydrogenation pathways qualitatively match the LP-TEM observations (\textbf{Figure 2} and \textbf{3}). At low or high $\beta$-phase fractions, the $\alpha$/$\beta$-$\mathrm{PdH}_x$ interface is significantly curved, most commonly near a nanocrystal corner (\textbf{Figure 4c} and \textbf{S17}, and \textbf{Movies S5–S7}). The interface then flattens as the system approaches $c_\beta$ $\sim$0.5, before curving again during relaxation (\textbf{Figure S17} and \textbf{Movies S5–S7}).

As a first step in quantitatively understanding the typical dynamics across the ensemble of all absorption trajectories, we calculate the average elastic energy as a function of the $\beta$-phase fraction (\textbf{Figure 4b}). For all chemical driving forces, the elastic energy increases as the $\beta$-phase fraction approaches 0.5, reflecting the strain that accumulates from the co-existence of the two $\mathrm{PdH}_x$ phases. In the absence of chemical driving ($\Delta$$\mu$ = 0.0), the elastic energy approximately plateaus over the range $c_\beta$ $\sim$0.3–0.7, then decreases in a manner symmetric to its initial rise. A statistical analysis of the crystallographic alignment of the $\alpha$/$\beta$-$\mathrm{PdH}_x$ interface (\textbf{Figure 4d} and see \textbf{Methods}) shows that within this plateau region, nearly all configurations in trajectories propagating under $\Delta$$\mu$ = 0.0 exhibit interfaces aligned along \{110\}.

Under a chemical driving force of $\Delta$$\mu$ = 0.2 $k_\mathrm{B}T$, the elastic energy initially exceeds the plateau value of the $\Delta$$\mu$ = 0.0 ensemble, then decreases beyond $c_\beta$ $\sim$0.4, ultimately following an energetically identical pathway to the $\Delta$$\mu$ = 0.0 case. \textbf{Figure 4d} reveals that this trend in the average elastic energy can be explained by a tilt from a \{211\} interfacial alignment toward a \{110\} alignment. This is qualitatively the same as the process observed in Nanoplate \#1 in the LP-TEM experiments (\textbf{Figure 3}, \textbf{4c}, and \textbf{S17}). In contrast, in the ensemble of trajectories driven more strongly ($\Delta$$\mu$ = 0.5 $k_\mathrm{B}T$), the elastic energy continues to rise up to $c_\beta$ $\sim$0.5. Analysis of the interfacial alignment (\textbf{Figure 4d}) shows that this pathway corresponds to one in which a \{211\} orientation is predominantly maintained throughout the trajectory. Typical trajectories in this ensemble closely resemble the LP-TEM observations of Nanoplate \#3 (\textbf{Figure 3}, \textbf{4c}, and \textbf{S17}).

Prior to $c_\beta$ $\sim$0.40, both driven hydrogenation ensembles ($\Delta$$\mu$ = 0.2 and 0.5 $k_\mathrm{B}T$) tend to favor a net interfacial alignment along \{211\} rather than \{110\} (\textbf{Figure 4d} and \textbf{S17}). Importantly, this $c_\beta$ range ($\lesssim 0.40$) also contains the free energy maxima for both driven ensembles: $c_\beta$ $\sim$0.39 for $\Delta$$\mu$ = 0.2 $k_\mathrm{B}T$ and $c_\beta$ $\sim$0.23 for $\Delta$$\mu$ = 0.5 $k_\mathrm{B}T$. At similar values of $c_\beta$ (between $\sim$0.20 and $\sim$0.36), the trajectories in the driven ensembles access configurations that are lower in elastic energy (\textbf{Figure 4b}) than those of the undriven ensemble ($\Delta$$\mu$ = 0.0). Thus, in all cases of chemical driving, the system appears to relax toward configurations with favorable elastic energies near the free energy maximum for that ensemble (\textbf{Figure 4a} and \textbf{4b}). However, in general, the system does not necessarily minimize the elastic energy during the relaxation process away from the free energy barrier.

When considering idealized, perfectly flat $\alpha$/$\beta$-$\mathrm{PdH}_x$ interfaces, there is a similar crossover in the elastic energy cost of interfacial alignments. An idealized \{211\} alignment becomes energetically favorable over an idealized \{110\} alignment in the $c_\beta$ range of approximately 0.10-0.24 (\textbf{Figure S15}). Because the interfaces in absorption trajectories are significantly curved and not perfectly aligned with either crystallographic orientation, the range over which an idealized \{211\} alignment is energetically favored will not exactly coincide with the range in which the driven ensembles have a lower average elastic energy than the undriven ensemble. However, the origin of these two energetic crossovers is the same. In the nanoplate geometry, a \{110\} interfacial alignment is energetically favored at $c_\beta$ = 0.5, but tilting toward \{211\} is favored at values of $c_\beta$ away from 0.5. This is why the location of the free energy barrier as a function of the $\beta$-phase fraction is crucially important for the determination of the hydrogenation pathway.

For $\Delta$$\mu$ = 0.2 $k_\mathrm{B}T$, the free energy maximum lies close to the point where the average elastic energy within both driven ensembles has exceeded the average within the $\Delta$$\mu$ = 0.0 ensemble. Beyond this point, for $c_\beta$ in the range 0.4–0.5, the absorption rate (see \textbf{Methods}) remains approximately constant (\textbf{Figure S16)} in the $\Delta$$\mu$ = 0.2 $k_\mathrm{B}T$ ensemble, and the average elastic energy decreases (\textbf{Figure 4b}). By contrast, for the $\Delta$$\mu$ = 0.5 $k_\mathrm{B}T$ ensemble, the free energy maximum occurs at a much lower $c_\beta$ ($\sim$0.23 vs. $\sim$0.39) and therefore well before the crossover in average elastic energy between the driven and undriven ensembles, which lies closer to $c_\beta$ $\sim$0.36 (\textbf{Figure 4b} and \textbf{4d}). The absorption rate in this ensemble rapidly accelerates in the range of $c_\beta$ = 0.4–0.5 (\textbf{Figure S16}), and the average elastic energy increases (\textbf{Figure 4b}). Thus, typical trajectories in the more slowly absorbing $\Delta$$\mu$ = 0.2 $k_\mathrm{B}T$ ensemble tilt toward the \{110\} configurations that are lower in elastic energy at $c_\beta$ = 0.5 (\textbf{Figure S15}), whereas trajectories in the $\Delta$$\mu$ = 0.5 $k_\mathrm{B}T$ ensemble maintain a \{211\} alignment throughout the absorption process, even at the expense of higher elastic energy (\textbf{Figure 4b}, \textbf{4d}, \textbf{S15}, and \textbf{S17}). In other words, our observations suggest that during such accelerated absorption, far from the free energy maximum, the system cannot relax to the energetically preferred configuration.

\subsection*{Theoretical modeling of phase transformation pathways in nanocubes}

\begin{figure*}[t]

\centering

\includegraphics[width=0.70\textwidth]{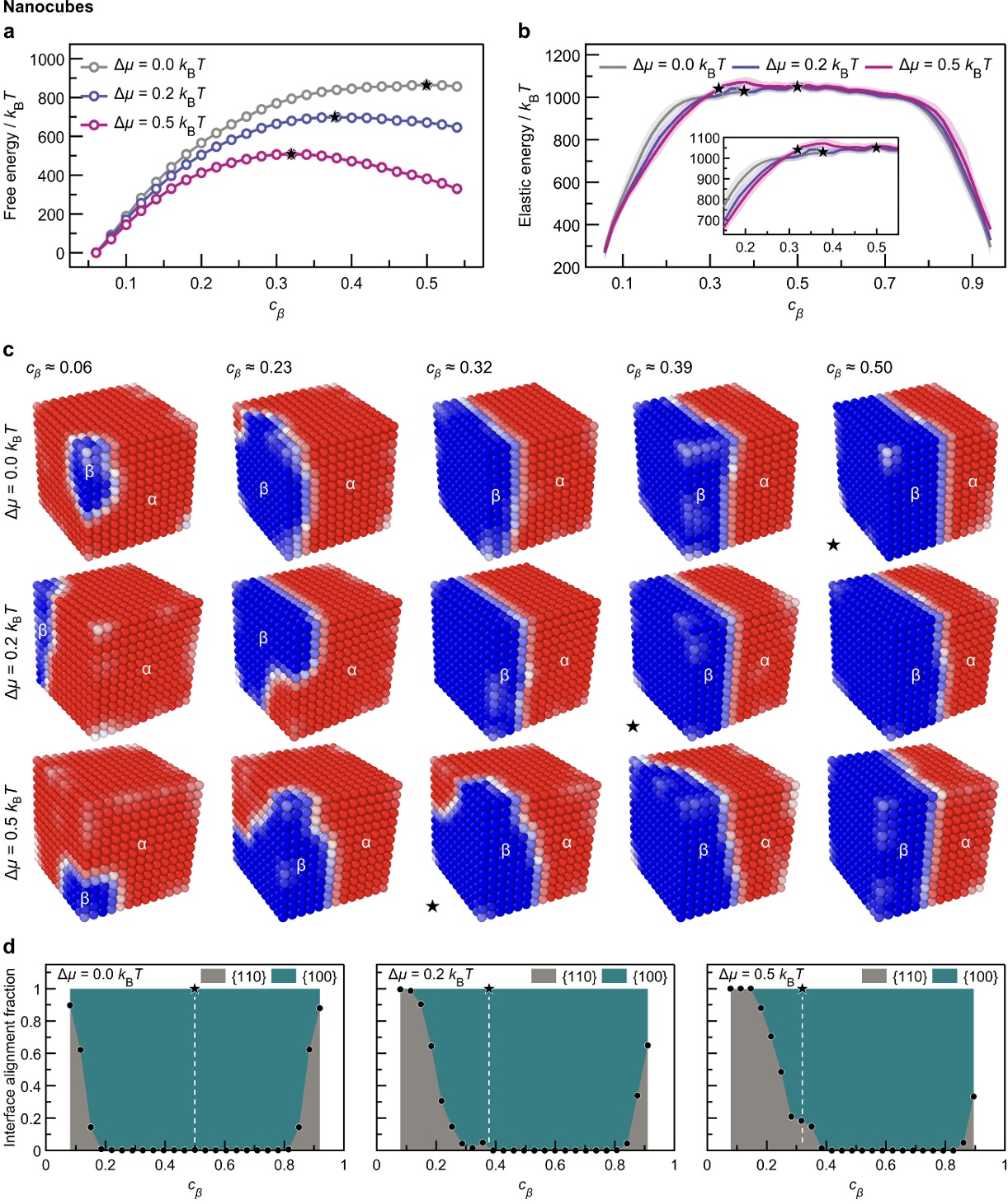}

\caption{Free energetics and simulated hydrogenation dynamics in nanocubes. \textbf{a}, Helmholtz free energy as a function of the $\beta$-phase fraction ($c_\beta$) in the nanocube. Black pentagram markers denote the transition state locations for each value of the effective chemical driving, $\Delta$$\mu$. \textbf{b}, Average elastic energy across simulated hydrogenation trajectories, plotted as a function of $c_\beta$. The shaded area represents the standard deviation across trajectories. Black pentagram markers correspond to the same transition-state $c_\beta$ values shown in panel \textbf{a} for each curve. All energies are measured in units of $k_\mathrm{B}T$, where $k_\mathrm{B}$ is the Boltzmann constant and $T$ is the simulation temperature. \textbf{c}, Representative snapshots from absorption trajectories in nanocubes. Each row corresponds to a single trajectory at $\Delta$$\mu$ = 0.0, 0.2, and 0.5 $k_\mathrm{B}T$ (top to bottom). Snapshots are colored in terms of the density of each phase, with blue as $\beta$-$\mathrm{PdH}_x$ and red as $\alpha$-$\mathrm{PdH}_x$. Black pentagram markers additionally indicate the snapshots corresponding to the transition state. \textbf{d}, Fraction of $\alpha$/$\beta$-$\mathrm{PdH}_x$ interfaces aligned with either the \{110\} or \{100\} orientations across the ensemble of simulated absorption trajectories. At each value of $c_\beta$, the $\alpha$/$\beta$ interface is computed for all trajectories. The color represents the fraction of best-fit planes with a normal vector closer to that crystallographic orientation than to the alternative orientation considered. Black pentagram markers and their corresponding vertical lines denote the transition-state $c_\beta$ values.}

\label{fig:5}

\end{figure*}

The nanoplate analysis reveals that chemical driving of sufficient magnitude enables access to a qualitatively different phase transformation pathway. We now examine the geometry dependence of this phenomenon by applying the same analysis to a nanocube of virtually identical size (2,197 sites vs. 2,133 sites in the nanoplate). We first note that, due to differences in both the free energy barrier height and the curvature of the barrier, the free energy maxima corresponding to each tested non-zero value of $\Delta$$\mu$ lie at different values of $c_\beta$ in the nanocube compared with the nanoplate (\textbf{Figure 5a} vs. \textbf{4a}).

Examining the average elastic energy along absorption paths (\textbf{Figure 5b}), one again observes that for $\Delta$$\mu$ = 0.0, there is an approximate plateau symmetric about $c_\beta$ = 0.5. However, this plateau spans a broader range of $c_\beta$ (approximately 0.2–0.8), and the $\alpha$/$\beta$-$\mathrm{PdH}_x$ interface with the lowest elastic energy at $c_\beta$ = 0.5 is aligned not along \{110\} but along \{100\} (\textbf{Figure 5c}, \textbf{5d}, and \textbf{S18}). Additionally, in both driven ensembles ($\Delta$$\mu$ = 0.2 and 0.5 $k_\mathrm{B}T$), the average elastic energy transiently exceeds the plateau of the undriven ensemble (\textbf{Figure 5b}), mirroring the behavior observed in the nanoplate. After $c_\beta$ = 0.5, however, both driven ensembles in the nanocube follow a pathway that closely resembles the undriven ensemble in terms of their elastic energies.

In the nanocube, the free energy maxima corresponding to all values of $\Delta$$\mu$ fall within the $c_\beta$ range where \{100\} alignment is favorable (\textbf{Figure 5d} and \textbf{S18}). Under chemical driving, an increase in \{110\}-aligned configurations does appear at lower $c_\beta$ values prior to the free energy maxima (\textbf{Figure 5d}), and these \{110\} configurations accompany a decrease in the average elastic energy (\textbf{Figure 5b}). However, once the barrier is crossed, the interface always tilts back toward a predominant \{100\} orientation. This behavior is closely analogous to the $\Delta$$\mu$ = 0.2 $k_\mathrm{B}T$ ensemble of the nanoplates, where a transient appearance of \{211\} alignment is followed by a return to the \{110\} orientation that minimizes elastic energy at $c_\beta$ = 0.5 (\textbf{Figure S15}). In the $\Delta$$\mu$ = 0.5 $k_\mathrm{B}T$ ensemble of the nanocube, one does observe a notable acceleration of the absorption rate (\textbf{Figure S16}) after the transition state ($c_\beta$ $\sim$0.31-0.40), mirroring the $\Delta\mu$ = 0.5 $k_\mathrm{B}T$ ensemble of the nanoplate. During this range of $c_\beta$, the average elastic energy (\textbf{Figure 5b}) initially increases ($c_\beta \approx$ 0.30-0.37) before gradually decreasing over $c_\beta \approx$ 0.37-0.40, reaching the same value as the $\Delta\mu$ = 0.0 and 0.2 $k_\mathrm{B}T$ nanocube ensembles. Unlike in the nanoplate, however, this acceleration in the nanocube notably occurs after the interface has already begun to tilt toward the energetically preferred \{100\} alignment (\textbf{Figure 5c} and \textbf{5d}). Thus, rather than establishing a qualitatively different hydrogenation pathway—as was observed in the nanoplate—this combination of nanocrystal geometry and chemical driving force ultimately still favors the formation of the energetically preferred interfacial alignment at $c_\beta$ = 0.5. These observations are also fully consistent with the LP-TEM observations~\cite{lee_atomic_2024} (\textbf{Figure 3}, \textbf{5c}, \textbf{S18}, and \textbf{S19}, and \textbf{Movies S8–S10}).

\subsection*{The influence of geometry on accessibility of far-from-equilibrium hydrogenation pathways}

Our computational analyses demonstrate that the dominant absorption pathway generally depends on both the elastic energy cost of different configurations of the $\alpha$/$\beta$-$\mathrm{PdH}_x$ interface and the hydrogen absorption rate at each point along the pathway. Crucially, these factors are all governed by nanocrystal geometry. The elastic energy of different interfacial alignments varies with the $\beta$-phase fraction ($c_\beta$) (\textbf{Figure S15}, \textbf{4b–4d}, and \textbf{5b–5d}), and which alignment is most favorable is determined by the nanocrystal geometry. The hydrogen absorption rate at any value of $c_\beta$ depends on the entire system’s proximity to its free energy barrier and is therefore determined by both the nanocrystal geometry and the chemical potential of hydrogen in solution. A larger chemical potential difference between hydrogen in solution and hydrogen absorbed in the $\mathrm{PdH}_x$ nanocrystal shifts the free energy barrier to lower values of $c_\beta$, and—depending on the nanocrystal geometry—different characteristic shapes of the critical nucleus may then have favorable elastic energies. Once the system crosses the free energy barrier and the absorption rate accelerates, it need not relax to the configurations that are lowest in elastic energy, and thus the subsequent hydrogenation pathway can change qualitatively.

Specifically, we observed that for the same set of chemical driving forces, the dominant hydrogenation pathway in the nanoplate geometry undergoes a qualitative change as a function of the chemical driving force, whereas the pathway in the nanocube does not. At $c_\beta$ $\sim$0.5, the energetically favorable \{110\} alignment in the nanoplate is replaced almost exclusively by a \{211\} alignment when the chemical driving force becomes sufficiently large. In contrast, at $c_\beta$ $\sim$0.5 in the nanocube, the \{100\} alignment is always dominant. Therefore, we conclude that the LP-TEM observation of both \{110\}- and \{211\}-aligned $\alpha$/$\beta$-$\mathrm{PdH}_x$ interfaces at comparable $\alpha$/$\beta$ phase fractions indicates a mixture of thermodynamically and kinetically preferred pathways. For the nanocube geometry at the same level of chemical driving, the kinetic and thermodynamic pathways are qualitatively identical.

\section*{Conclusion}

Leveraging \textit{in situ} liquid phase TEM and theoretical modeling, we demonstrate that nanocrystal geometry can facilitate a diversity of phase transformation pathways in palladium hydride. We directly visualize hydrogenation in two model geometries—a Pd nanocube and hexagonal nanoplates—and reveal that, although both undergo comparable transformation sequences, their evolving $\alpha$/$\beta$-$\mathrm{PdH}_x$ interfaces exhibit geometry-dependent crystallographic orientations. In nanocubes, at the intermediate stages, the alignment along \{100\} planes reflects elastic energy ordering, whereas in nanoplates, the emergence of both \{110\}- and \{211\}-aligned interfaces—particularly the latter—indicates that geometry necessitates consideration of dynamic contributions beyond static energetics. Complementary theoretical simulations reveal that, under the same chemical driving force, the nanoplate geometry permits an alternative \{211\}-aligned pathway when the system is driven far from equilibrium, whereas the nanocube remains constrained to a \{100\}-aligned trajectory that maintains a local mechanical equilibrium.

These findings suggest that nanocrystal geometry can serve as a control parameter governing which phase transformation pathways are accessible during solid-state reactions involving solute intercalation. The observed coupling between geometry and transformation pathway provides a mechanistic framework that could be used to elucidate a broad range of solute-induced structural transformations across diverse material systems. Moreover, the insights from this study may introduce geometric control as a synthetic handle for directing materials along novel transformation pathways, thereby enabling access to configurations otherwise inaccessible under near-equilibrium conditions. Considering that phase-boundary morphologies influence stress accumulation and mechanical deformation during intercalation, our findings can also introduce new design principles through control of nanocrystal geometry, thereby enhancing the performance and durability of next-generation intercalation-mediated devices.

\section*{Methods}

\subsection*{Chemicals and materials}

Poly(vinyl pyrrolidone) (PVP, average molecular weight $\approx55,000$), L-ascorbic acid (AA, BioXtra, $99\%$), potassium bromide (KBr, 99.9\%), and an aqueous solution of KOH are purchased from Sigma-Aldrich. Sodium tetrachloropalladate(II) ($\mathrm{Na}_2\mathrm{PdCl}_4$) is obtained from Acros Organics for the nanocube synthesis and from Sigma-Aldrich for the nanoplate synthesis. All reagents are used as received without further purification. Aqueous solutions are prepared using deionized (DI) water with a resistivity of $18.2$  $\mathrm{M\Omega}\cdot \mathrm{cm}$ at room temperature. TEM grids coated with ultrathin carbon films (3–4 nm thick) are purchased from Electron Microscopy Sciences.

\subsection*{Pd nanocrystal synthesis}

The Pd nanocubes are synthesized following our reported procedure~\cite{jin_synthesis_2011}. In a typical synthesis, 4.0 mL of an aqueous solution containing 52.5 mg of PVP, 30 mg of AA, and 1.0 g of KBr is preheated to $80^\circ \mathrm{C}$ for 10 min under magnetic stirring. Subsequently, 1.5 mL of a separate aqueous solution comprising 28.5 mg of $\mathrm{Na}_2\mathrm{PdCl}_4$ is rapidly injected into the preheated mixture in a single step. The reaction is maintained at $80^\circ \mathrm{C}$ for 3 h. The resulting nanocubes are collected by centrifugation and washed three times with DI water.

The Pd nanoplates are synthesized based on our established protocol~\cite{xiong_polyvinyl_2006}. In a 25 mL three-neck flask equipped with a reflux condenser and a Teflon-coated magnetic stir bar, 105 mg of PVP is dissolved in 8.0 mL of DI water and heated to $90^\circ \mathrm{C}$ in air under magnetic stirring. Meanwhile, 3.0 mL of an aqueous solution of $\mathrm{Na}_2\mathrm{PdCl}_4$ (57 mg) is quickly added to the flask. The reaction is kept at  $90^\circ \mathrm{C}$ in air for 10 h. The resulting product is separated by centrifugation, washed once with acetone, and subsequently rinsed three times with ethanol to remove excess PVP. The AFM topographic mapping on the nanoplates is performed using a Park XE7 atomic force microscope (Park Systems).

\subsection*{Liquid cell assembly}

For \textit{in situ} LP-TEM observations, liquid cells are assembled from TEM grids coated with ultrathin carbon films (3–4 nm thick). The assembly process involves encapsulating a thin layer of aqueous KOH solution and Pd nanocrystals between two carbon films through van der Waals adhesion (schematically illustrated in \textbf{Figure S6}). Each TEM grid is plasma-cleaned using a Model 1020 Plasma Cleaner (Fischione Instruments) with a mixed argon–oxygen atmosphere to render the carbon-film surface hydrophilic and to remove residual organic contaminants. A droplet (1–2 $\mu$L) of Pd nanocrystal dispersion is drop-cast onto the plasma-treated grid and allowed to dry in air. A second pristine TEM grid coated with the same carbon film is plasma-cleaned under identical conditions. A small droplet ($\leq$ 0.5 $\mu$L) of aqueous KOH solution is then applied to the nanocrystal-loaded grid; KOH concentrations of 0.005–0.01 M are used. The second TEM grid is gently placed atop the first, encapsulating the Pd nanocrystals and KOH solution to form a carbon-film-based liquid cell. The assembled cell is finally loaded into the transmission electron microscope, where the hydrogenation of the Pd nanocrystals is directly visualized \textit{in situ}.

\subsection*{Transmission electron microscopy}

\textit{In situ} LP-TEM, together with \textit{ex situ} HRTEM, SAED, and STEM-EDS investigations, is carried out on a FEI ThemIS 60-300 transmission electron microscope (Thermo Fisher Scientific). The microscope, operated at 300 kV, is equipped with a Ceta2 complementary metal-oxide semiconductor (CMOS) camera coupled with a scintillator, an image aberration corrector capable of achieving 70 pm spatial resolution, and a SuperX EDS detector consisting of four windowless silicon drift detectors.

\textit{Ex situ} (HR)STEM imaging is performed using the Transmission Electron Aberration-corrected Microscope I (TEAM I), a modified FEI Titan 80-300 microscope, at the National Center for Electron Microscopy, Molecular Foundry, Lawrence Berkeley National Laboratory, operated at 300 kV. The instrument is equipped with a CEOS DCOR spherical-aberration STEM probe corrector and a high-angle annular dark-field (HAADF)-STEM detector. Post-acquisition correction of drift-induced distortions in HRSTEM images is performed using image pairs acquired with orthogonal scan directions and code in the open-source package \textit{quantem}, developed based on previous approaches\cite{ophus_correcting_2016}(\textbf{Figure S2}).

\subsection*{HRTEM image analysis}

The \textit{d}-spacing colormaps are generated from the corresponding HRTEM images using the GPA method~\cite{hytch_analysis_1997,hytch_quantitative_1998}. In this approach, a pair of Bragg reflections is selected in the Fourier domain, and one reflection is designated as the reference reciprocal lattice vector ($\mathbf{g}$). A two-dimensional Bragg filter centered on this reflection is constructed using a flat-topped circular mask with a gradual intensity decay toward zero near its periphery~\cite{lee_atomic_2024}. The Fourier transform of the HRTEM image is multiplied by this Bragg-filtering mask, and the inverse Fourier transform of the Bragg-filtered Fourier transform yields a complex image, $H'_g(\mathbf{r})$, where $\mathbf{r}$ denotes the spatial coordinates in real space. The amplitude image, $A_g(\mathbf{r})$, is obtained as the modulus of $H'_g(\mathbf{r})$, while the corresponding phase image, $P_g(\mathbf{r})$, is derived from its argument after removing the term $2\pi\mathbf{g}\cdot\mathbf{r}$. The strain map, $\epsilon_g(\mathbf{r})$, is then determined from the spatial derivatives of $P_g(\mathbf{r})$ relative to $\mathbf{g}$ and subsequently converted into a \textit{d}-spacing map, $d_g(\mathbf{r})$, using the \textit{d}-spacing corresponding to the reference reciprocal vector.

To account for the contributions of all symmetry-equivalent reflections, \textit{d}-spacing maps from individual reflections are combined through amplitude weighting of the corresponding Bragg-filter outputs (see \textbf{Figure S10}). Specifically, the \{200\} family is constructed from the (200) and (020) reflections, the \{220\} family from the (022), (202), and (2-20) reflections, and the $\frac{1}{3}$\{422\} family from $\frac{1}{3}$(-242), $\frac{1}{3}$(224), and $\frac{1}{3}$(4-22) reflections. The final amplitude-weighted \textit{d}-spacing colormap, $d_{hkl}(\mathbf{r})$, is calculated as $d_{hkl}(\mathbf{r}) = \sum_i A_{gi}(\mathbf{r})^2 d_{gi}(\mathbf{r}) / \sum_i A_{gi}(\mathbf{r})^2$, where $A_{gi}(\mathbf{r})$ represents the amplitude image of each reflection within the crystallographic family and $d_{gi}(\mathbf{r})$ the corresponding $d$-spacing map.

\subsection*{Computational methods}

We perform all simulations using the coarse-grained model based on the elastic Ising model~\cite{fratzl_modeling_1999,frechette_consequences_2019,frechette_elastic_2021}, and parameterized~\cite{lee_atomic_2024} for the $\mathrm{PdH}_x$ system. In the model, the total system energy can be written as
$$E_\mathrm{total}= E_\mathrm{elastic}-\Delta\mu c_\beta N = \frac{1}{2}\sum_{\mathbf{R},\mathbf{R'}}\sigma_\mathbf{R}V_{\mathbf{R},\mathbf{R'}}\sigma_{\mathbf{R'}} -\Delta\mu c_\beta N$$

where $\Delta$$\mu$ is the chemical potential difference between hydrogen in solution and in the $\beta$-$\mathrm{PdH}_x$ phase, $N$ is the number of unit cells in the system, and $c_\beta$ is the $\beta$-phase fraction. Each unit cell is represented by a single site corresponding to a position in the nanocrystal, R, with a spin, $\sigma_\mathbf{R} = \pm 1$ corresponding to the $\beta$ and $\alpha$ phases, respectively. The term $V_{\mathbf{R},\mathbf{R'}}$ in the elastic energy is an effective interaction between sites $\mathbf{R}$ and $\mathbf{R'}$, which depends on the nanocrystal geometry and replicates the elastic response of palladium~\cite{lee_atomic_2024}. The sites are represented on a simple cubic lattice in both nanocrystal geometries.

For all simulations, we propagate the dynamics using the standard Gillespie algorithm~\cite{gillespie_stochastic_2007}. The transition rate from configuration $\mathbf{C}$ to $\mathbf{C'}$ takes the form,
$$k_{\mathbf{C},\mathbf{C'}}=k_0\exp\bigg(\frac{-\big(E_\mathrm{total}(\mathbf{C'}) - E_\mathrm{total}(\mathbf{C})\big)}{2k_\mathrm{B}T}\bigg)$$

Here, $E_{\mathrm{total}}(\mathbf{C})$ denotes the total energy of configuration $\mathbf{C}$, $k_0$ is the pre-factor being constant for all processes, which defines the simulation time unit $\tau$ = $k_0^{-1}$, $k_{\mathrm{B}}$ is Boltzmann’s constant, and T is the simulation temperature, which we set to 100 K for all simulations. Note that this simulation temperature cannot be directly related to the experimental temperature because the nanocrystal sizes in the simulations are significantly smaller than those in the experiments, and the critical temperature of the model is strongly size-dependent. Using 100 K ensures that the interface can be quantified precisely without excessive statistical noise; based on previous calculations at 300 K~\cite{lee_atomic_2024}, we expect our results to be qualitatively insensitive to temperature, so long as the system remains well below its size-dependent critical temperature. This condition is satisfied in both experiment and simulation.

The allowed moves in kinetic Monte Carlo represent absorption/desorption and diffusion. A spin can flip if it lies on the boundary of the nanocrystal (i.e., the site has fewer than six nearest neighbors). Any two nearest-neighbor sites with opposite spins can exchange their spin values.

The free energies shown in \textbf{Figure 4a} and \textbf{5a} are calculated via umbrella sampling using the Weighted Histogram Analysis Method~\cite{grossfield_wham}. For a sampling window centered at $\beta$-phase fraction $c_\beta$*, an umbrella potential is added to the total energy of the form $U=10Nk_\mathrm{B}T(c_\beta-c_\beta^*)^2/2$. For each system, we use 512 windows with centers between $c_\beta = -0.06$ and 0.55.

To generate absorption trajectories, we propagate dynamics forward and backward in time, starting from the top of the free energy barrier~\cite{bolhuis_transition_2002}. In the case of kinetic Monte Carlo dynamics, time reversal is trivial as there is no momentum. Thus, a configuration is selected at random from umbrella sampling trajectories near the top of the free energy barrier. Then, two trajectories are propagated from that same configuration in the absence of the umbrella potential. If one trajectory reaches $c_\beta > 0.95$ and the other reaches $c_\beta < 0.05$, we reverse the latter in time and concatenate the two trajectories to form an absorption pathway. If this does not occur, the attempt to generate an absorption pathway is rejected, and a new attempt begins. For each condition tested in each geometry, we generate at least 62 independent absorption trajectories in this fashion.

Note that this method of generating the absorption pathway ensemble assumes that $c_\beta$ is the true reaction coordinate in that the set of configurations sampled from the free energy maximum is a representative sample of the transition state ensemble. Although this may not be strictly exact, any absorption event must increase $c_\beta$, so the true reaction coordinate must be strongly correlated with $c_\beta$. Therefore, we expect that our qualitative picture of the absorption-path ensemble remains accurate.

To produce the density maps in \textbf{Figure 4c}, \textbf{5c}, and \textbf{S17–S19}, and \textbf{Movies S5–S10}, we follow the procedure of Willard and Chandler~\cite{willard_instantaneous_2010}, convoluting the discrete density field with Gaussian distributions, such that the density for a given site is equal to,

$$\tilde{c}_\beta(\mathbf{R}) = \frac{1}{2\mathcal{N}}\sum_{\mathbf{R'}}(\sigma_{\mathbf{R'}}+1)e^{-|\mathbf{R'}-\mathbf{R}|^2/2a^2}$$

where a is the lattice constant of the model. In practice, the sum is cut off at a distance of $2a$, and $\mathcal{N}$ is the normalization of the discretized Gaussian over this range.

For all configurations in the simulated absorption trajectories, we identify the interfacial sites by extracting the boundary of the largest contiguous cluster of sites with  $\tilde{c}_\beta< 0.5$. Once $M$ interfacial sites are identified, we then determine the normal vector of the best-fit plane to the interface via singular value decomposition of the $M \times 3$ matrix of interfacial site positions. The crystallographic orientation with the largest projection onto this normal vector is defined as the interfacial alignment. To remain consistent with the imaging capabilities of LP-TEM, we focus only on the \{110\} and \{211\} orientations in the nanoplate geometry, and the \{110\} and \{100\} orientations in the nanocube geometry.

\section*{Data And Code Availability}

The data needed to evaluate the conclusions in this work are present in the paper and Supporting Information are available from the corresponding authors upon reasonable request. The codes used for the analysis and simulation are available from the lead contact upon request.

\section*{Acknowledgments}

This work was supported by the U.S. Department of Energy, Office of Science, Office of Basic Energy Sciences (BES), Materials Sciences and Engineering Division under Contract No. DE-AC02-05-CH11231 within the \textit{in situ} TEM program (KC22ZH). Work at the Molecular Foundry at Lawrence Berkeley National Laboratory was supported by the U.S. Department of Energy under Contract No. DE-AC02-05CH11231. D.L. acknowledges the Kwanjeong Study Abroad Scholarship from the KEF (Kwanjeong Educational Foundation) (KEF-2019). H.M. and J.Y. were financially supported by the Ministry of Trade, Industry and Energy (MOTIE) and the Korea Institute for Advancement of Technology (KIAT) through the International Cooperative R\&D program (P0026257).

\section*{Author Contributions}

D.L. and H.M. performed the \textit{in situ} liquid phase TEM and \textit{ex situ} S/TEM experiments under the supervision of H.Z. J.H., Z.W., and Y.S. synthesized the Pd nanocrystals under the supervision of Y.X. D.L., H.M., S.R., and C.O. analyzed the experimental S/TEM data. E.A. and H.M. conducted AFM measurements under the supervision of J.Y. K.C.B., C.S., and R.D. assisted with both \textit{in situ} and \textit{ex situ} S/TEM experiments. S.O.-L. carried out the theoretical modeling and simulations and interpreted the results under the supervision of D.T.L. The manuscript was written by D.L., S.O.-L., D.T.L., and H.Z., with input from all co-authors. H.Z. conceived the study and supervised the project.

\section*{Conflict Of Interest}

The authors declare no competing interests.

TABLE OF CONTENTS (TOC)

Nanocrystal geometry governs phase transformation pathways in palladium hydride ($\mathrm{PdH}_x$). \textit{in situ} liquid phase transmission electron microscopy reveals that nanocubes follow elastic-energy-ordered \{100\} $\alpha$/$\beta$-$\mathrm{PdH}_x$ interfacial alignments, whereas nanoplates access both \{110\} and \{211\} configurations indicative of an alternative route. Theoretical simulations confirm geometry as a control knob enabling atypical solute-mediated transformation pathways, such as \{211\}, which manifest under far-from-equilibrium conditions.

\section*{References}
\bibliography{main.bib}

\end{document}